\documentclass{webofc}
\usepackage[varg]{txfonts}   % Web of Conferences font
\begin{document}
\title{The NIKA polarimeter on science targets}
\subtitle{Crab nebula observations at 150 GHz and dual-band polarization images of Orion Molecular Cloud OMC-1}

\author{\firstname{A.} \lastname{Ritacco} \inst{\ref{IAS},\ref{IRAME}} \fnsep\thanks{\email{alessia.ritacco@ias.u-psud.fr}}
\and \firstname{R.} \lastname{Adam} \inst{\ref{LLR},\ref{CEFCA}}
\and  \firstname{P.} \lastname{Ade} \inst{\ref{Cardiff}}
\and  \firstname{P.} \lastname{Andr\'e} \inst{\ref{CEA1}}
\and  \firstname{A.} \lastname{Andrianasolo} \inst{\ref{IPAG}}
\and  \firstname{H.} \lastname{Aussel} \inst{\ref{CEA1}}
\and  \firstname{A.} \lastname{Beelen} \inst{\ref{IAS}}
\and  \firstname{A.} \lastname{Beno\^it} \inst{\ref{Neel}}
\and  \firstname{A.} \lastname{Bideaud} \inst{\ref{Neel}}
\and  \firstname{O.} \lastname{Bourrion} \inst{\ref{LPSC}}
\and  \firstname{M.} \lastname{Calvo} \inst{\ref{Neel}}
\and  \firstname{A.} \lastname{Catalano} \inst{\ref{LPSC}}
\and  \firstname{B.} \lastname{Comis} \inst{\ref{LPSC}}
\and  \firstname{M.} \lastname{De~Petris} \inst{\ref{Roma}}
\and  \firstname{F.-X.} \lastname{D\'esert} \inst{\ref{IPAG}}
\and  \firstname{S.} \lastname{Doyle} \inst{\ref{Cardiff}}
\and  \firstname{E.~F.~C.} \lastname{Driessen} \inst{\ref{IRAMF}}
\and  \firstname{A.} \lastname{Gomez} \inst{\ref{CAB}}
\and  \firstname{J.} \lastname{Goupy} \inst{\ref{Neel}}
\and  \firstname{F.} \lastname{K\'eruzor\'e} \inst{\ref{LPSC}}
\and  \firstname{C.} \lastname{Kramer} \inst{\ref{IRAME}}
\and  \firstname{B.} \lastname{Ladjelate} \inst{\ref{IRAME}}
\and  \firstname{G.} \lastname{Lagache} \inst{\ref{LAM}}
\and  \firstname{S.} \lastname{Leclercq} \inst{\ref{IRAMF}}
\and  \firstname{J.-F.} \lastname{Lestrade} \inst{\ref{LERMA}}
\and  \firstname{J.F.} \lastname{Mac\'ias-P\'erez} \inst{\ref{LPSC}}
\and  \firstname{P.} \lastname{Mauskopf} \inst{\ref{Cardiff},\ref{Arizona}}
\and \firstname{A.} \lastname{Maury} \inst{\ref{CEA1}}
\and \firstname{F.} \lastname{Mayet} \inst{\ref{LPSC}}
\and  \firstname{A.} \lastname{Monfardini} \inst{\ref{Neel}}
\and  \firstname{L.} \lastname{Perotto} \inst{\ref{LPSC}}
\and  \firstname{G.} \lastname{Pisano} \inst{\ref{Cardiff}}
\and  \firstname{N.} \lastname{Ponthieu} \inst{\ref{IPAG}}
\and  \firstname{V.} \lastname{Rev\'eret} \inst{\ref{CEA1}}
\and  \firstname{C.} \lastname{Romero} \inst{\ref{IRAMF}}
\and  \firstname{H.} \lastname{Roussel} \inst{\ref{IAP}}
\and  \firstname{F.} \lastname{Ruppin} \inst{\ref{MIT}}
\and  \firstname{K.} \lastname{Schuster} \inst{\ref{IRAMF}}
\and  \firstname{S.} \lastname{Shu} \inst{\ref{IRAMF}}
\and  \firstname{A.} \lastname{Sievers} \inst{\ref{IRAME}}
\and  \firstname{C.} \lastname{Tucker} \inst{\ref{Cardiff}}
\and  \firstname{R.} \lastname{Zylka} \inst{\ref{IRAMF}}}

\institute{\label{LPSC} Univ. Grenoble Alpes, CNRS, Grenoble INP, LPSC-IN2P3, 53, avenue des Martyrs, 38000 Grenoble, France
\and \label{LLR} LLR (Laboratoire Leprince-Ringuet), CNRS, \'Ecole Polytechnique, Institut Polytechnique de Paris, Palaiseau, France  
\and \label{CEFCA} Centro de Estudios de F\'isica del Cosmos de Arag\'on (CEFCA), Plaza San Juan, 1, planta 2, E-44001, Teruel, Spain 
\and \label{Cardiff} Astronomy Instrumentation Group, University of Cardiff, UK          
\and \label{CEA1} AIM, CEA, CNRS, Universit\'e Paris-Saclay, Universit\'e Paris Diderot, Sorbonne Paris Cit\'e, 91191 Gif-sur-Yvette, France     
\and \label{IPAG} Univ. Grenoble Alpes, CNRS, IPAG, 38000 Grenoble, France     
\and \label{IAS} Institut d'Astrophysique Spatiale (IAS), CNRS and Universit\'e Paris Sud, Orsay, France    
\and \label{Neel} Institut N\'eel, CNRS and Universit\'e Grenoble Alpes, France
\and \label{Roma} Dipartimento di Fisica, Sapienza Universit\`a di Roma, Piazzale Aldo Moro 5, I-00185 Roma, Italy       
\and \label{IRAMF} Institut de RadioAstronomie Millim\'etrique (IRAM), Grenoble, France 
\and \label{CAB} Centro de Astrobiolog\'ia (CSIC-INTA), Torrej\'on de Ardoz, 28850 Madrid, Spain
\and \label{IRAME} Instituto de Radioastronom\'ia Milim\'etrica (IRAM), Granada, Spain 
\and \label{LAM} Aix Marseille Univ, CNRS, CNES, LAM (Laboratoire d'Astrophysique de Marseille), Marseille, France
\and \label{LERMA} LERMA, Observatoire de Paris, PSL Research University, CNRS, Sorbonne Universit\'es, UPMC Univ. Paris 06, 75014 Paris,
France
\and \label{Arizona} School of Earth and Space Exploration and Department of Physics, Arizona State University, Tempe, AZ 85287         
\and \label{IAP} Institut d'Astrophysique de Paris, CNRS (UMR7095), 98 bis boulevard Arago, 75014 Paris, France
\and \label{MIT} Kavli Institute for Astrophysics and Space Research, Massachusetts Institute of Technology, Cambridge, MA 02139, USA 
          }

\newcommand{\nika}{{\it NIKA}}
\newcommand{\nikad}{{\it NIKA2}}

\abstract{We present here the polarization system of the NIKA camera and give a summary of the main results obtained and performed studies on Orion and the Crab nebula. The polarization system was equipped with a room temperature continuously rotating multi-mesh half wave plate and a grid polarizer facing the NIKA cryostat window.
  NIKA even though less sensitive than NIKA2 had polarization capability in both 1 and 2 millimiter bands. NIKA polarization observations demonstrated the ability of such a technology in detecting the polarization of different targets, compact and extended sources like the Crab nebula and Orion Molecular Cloud region OMC-1.
   These measurements  together with the developed techniques to deal with systematics, opened the way to the current observations of NIKA2 in polarization that will provide important advances in the studies of galactic and extra-galactic emission and magnetic fields.
}
\maketitle
\section{Introduction}
\label{intro}
%The NIKA camera based at the IRAM 30m telescope from 2012 to 2015, has addressed different scientific topics that opened the way to current and future NIKA2 observations.
High angular resolution observations provided by the NIKA/NIKA2 cameras together with all sky maps provided by satellites, e.g. Planck and/or Herschel, are the key to understand a large number of physical processes from galactic to cosmological scales.
High angular resolution measurements in polarization are needed to disentangle the dependency of the polarization fraction on the magnetic field structure which has been traced by Planck satellite at large scales.
Planck \cite{plancki2014} and Herschel \cite{molinari2010, arzo2011} satellites revealed large scale filamentary structures as preferential sites of star formation.
These filamentary structures are associated with organized magnetic fields topology at scales larger than 0.5 pc \cite{andre2014} and indicate that magnetic field must be explored at scales of 0.01-0.1 pc \cite{pereyra2004, planckxxxiii2016}. The missing information at these scales will be provided by polarization observations with the NIKA2 camera.

 Polarization continuum observations performed with NIKA at 1.15 and 2.05 mm during the design stage of the NIKA2 polarization system have provided important information on the physics of two well known targets for polarization measurements: Orion Molecular Cloud OMC-1 and the Crab nebula. 
 %OMC-1 represents a typical astrophysical target for future NIKA2Pol observations, aiming at tracing magnetic fields structures in star forming regions.
 Thanks to the dual-band capability of NIKA we could trace the spatial distribution of the spectral index in polarization of Orion OMC-1 which showed interesting results along the bar, in the southern part of the source. The polarized emission of the main filament is detected and it shows a well ordered magnetic field following the total intensity structure of the source \cite{ritacco2017}.
The Crab nebula is a supernova remnant exhibiting a highly polarized synchrotron radiation at radio and millimeter wavelengths \cite{hester2008}. It is the brightest polarization source in the microwave sky with an extension of 7 by 5 arcminutes and commonly used as a standard candle for any experiment which aims at measuring the polarization of the sky \cite{macias2010,aumont2010}. Using the NIKA high resolution observation at 150 GHz and the observations of CMB satellites, WMAP and Planck, we could trace for the first time the Spectral Energy Distribution in polarization in the frequency range: 30-353 GHz. In this short paper we give an overview of the most interesting results obtained on these two targets.

%The NIKA pathfinder is a dual-band camera consisting of two arrayys filled by Lumped Elements Kinetic Inductance Detectors (LEKIDs) with a Hilbert geometry (Roesch et al. 2012).

\section{NIKA camera}
\label{nikapolsys}
The NIKA pathfinder is a dual-band KIDs (Kinetic Inductance Detectors) camera that observed the sky in two millimeter bands from the IRAM 30m telescope located in Pico Veleta, Spain.
With 132 detectors at 1.15 mm and 224 at 2.05 mm the NIKA camera covered 1.8' of the telescope field-of-view. The angular resolution was of 12 and 18.2 arcsec at 1.15 and 2.05 mm respectively \cite{monfardini2011}.
The NIKA calibration and performances are widely described in \cite{monfardini2010,monfardini2011,Catalano:2014nml}. Besides several scientific campaigns for total power observations, NIKA has been used also as test bench for the definition of the polarization channel of NIKA2 \cite{Calvo2016,NIKA2-Commissioning,NIKA2-Adam}. The NIKA2 camera with an increased number of detectors ($\sim$ 2900), field-of-view (6.5') and sensitivity is the current continuum camera of the IRAM 30m telescope.

\subsection{Polarization system}
\label{polsyst}
The NIKA polarization system consists of a continuously rotating half-wave-plate (HWP) and a grid polarizer placed at ambient temperature in front of the NIKA cryostat \cite{ritacco2016,ritacco2017}.
The polarizer is needed because of the NIKA detectors geometry \cite{roesch} that makes them not intrinsically sensitive to the polarization orientation.

The continuously rotating HWP allows to better filter the signal, and it allows the instantaneaous and simultaneous measurements of I, Q, and U\cite{ritacco2017}. This technique is possible thanks to the small time constant of the NIKA detectors of the order of 0.1 ms. Expansive description of the detection strategy, the technique to reconstruct the polarization from a modulated signal, the systematic effects, the calibration and performance of such a system can be found in \cite{ritacco2017}. The developed techniques and the dedicated data analysis software developed to provide Stokes I, Q, and U maps are currently used for NIKA2 polarization observations. The preliminary results of the NIKA2 commissioning phase will be presented in a companion paper.
The NIKA sensitivity in polarization mode corresponds to NEFD of 120 and 50 mJy s$^{1/2}$ at 1.15 and 2.05 mm, respectively. 

%Don't forget to give each section, subsection, subsubsection, and
%paragraph a unique label (see Sect.~\ref{sec-1}).

\section{Dual band observations of Orion Molecular Cloud OMC-1}
During the observational campaign of February 2015, the last one with the NIKA camera at the IRAM 30m telescope, we could observe the Orion star forming region OMC-1. This is the closest site of OB star formation. The nebula KL is located at the flux peak from far infrared to millimeter wavelengths on the OMC-1 ridge \cite{schleuning98}. The polarization maps and deeper analysis on the NIKA results can be found in \cite{ritacco2017}. Figure~1 shows the Stokes I maps obtained with the NIKA camera at both 260 GHz (left panel) and 150 GHz (right panel). Polarization vectors are overplotted, showing the polarization fraction and the angle. To first approximation we can consider the projected magnetic field component on the plane of the sky perpendicular to the direction of polarization, we observe a very organized magnetic field topology with field lines mostly orientated parallel to the integral-shaped filament. The polarization fraction reaches a level of 10$\%$ of the total intensity along the main filament where the diffuse emission is observed. It decreases greatly near the KL nebula position, corresponding to the peak of the total intensity, to a value of 0.6$\pm$0.2$\%$. The peak surface brightness is about 45.8 Jy/beam and 14 Jy/beam at 260 GHz and 150 GHz, respectively. \cite{ritacco2017} show that polarization is only detected at column densities NH$_{2}$ $>3$ x 10$^{23}$ cm$^{-2}$ in the map. Depolarization is observed at column densities NH$_{2}$ $>$ 4 x 10$^{24}$ cm$^{-2}$ at 1.15 mm while at 2.05 mm polarized fluxes seem to be less sensitive to depolarization, see the right panel of Figure~2 for comparison.

Thanks to the dual band capability of NIKA we could trace the spatial spectral index distribution in both total intensity and polarization. The emission of Orion OMC-1 is expected to be described by a thermal dust emission spectrum I$_{\nu}$ = I$_0\nu^{\beta_d}B_{\nu}(T_d)$, where $\beta_d$ is the dust spectral index, T$_d$ the dust temperature and B$_{\nu}$(T$_d$) the Planck spectrum. At NIKA frequencies the Planck spectrum B$_{\nu}$ reduces to Rayleigh-Jeans law: B$_{\nu}(T_d)$ = $\frac{2\nu^2}{c^2}$k$_B$T$_{RJ}$. As a consequence I$_{\nu}$ = I$_{0}\frac{2}{c^2}k_BT_{RJ}\nu^{\beta_d+2}$. Writing $\beta^{\prime}$ = $\beta_{d}$+2 we can compute $\beta^{\prime}$ = log($\frac{I_{\nu1}}{I_{\nu2}}$)/log($\frac{\nu1}{\nu2}$).

Figure~2 shows the spectral index $\beta^{\prime}$ maps obtained degrading the 1 mm map to the resolution of the 2 mm channel corresponding to 18.2 arcsec. Along the Orion bar in the southern part of the source the total intensity emission at 2 mm is prominent w.r.t the 1 mm map. This effect is translated into a drastic change of the spectral index that could be indicative of a change in the dust grain properties. Along the ridge, where the polarization is widely detected, we observe a spectral index $\beta^{\prime}$ that varies from 2 to 4 which is consistent with dust emission but for left side along the filament in total intensity where again the strong emission at 2 mm changes drastically the spectral index values to 0.25-1. This deserves a deeper investigation using multi-wavelenght information.

%For one-column wide figures use syntax of figure~\ref{fig-1}

\begin{figure}[ht]
  \begin{center}
   % \vspace*{5cm}
    \includegraphics[scale=0.25]{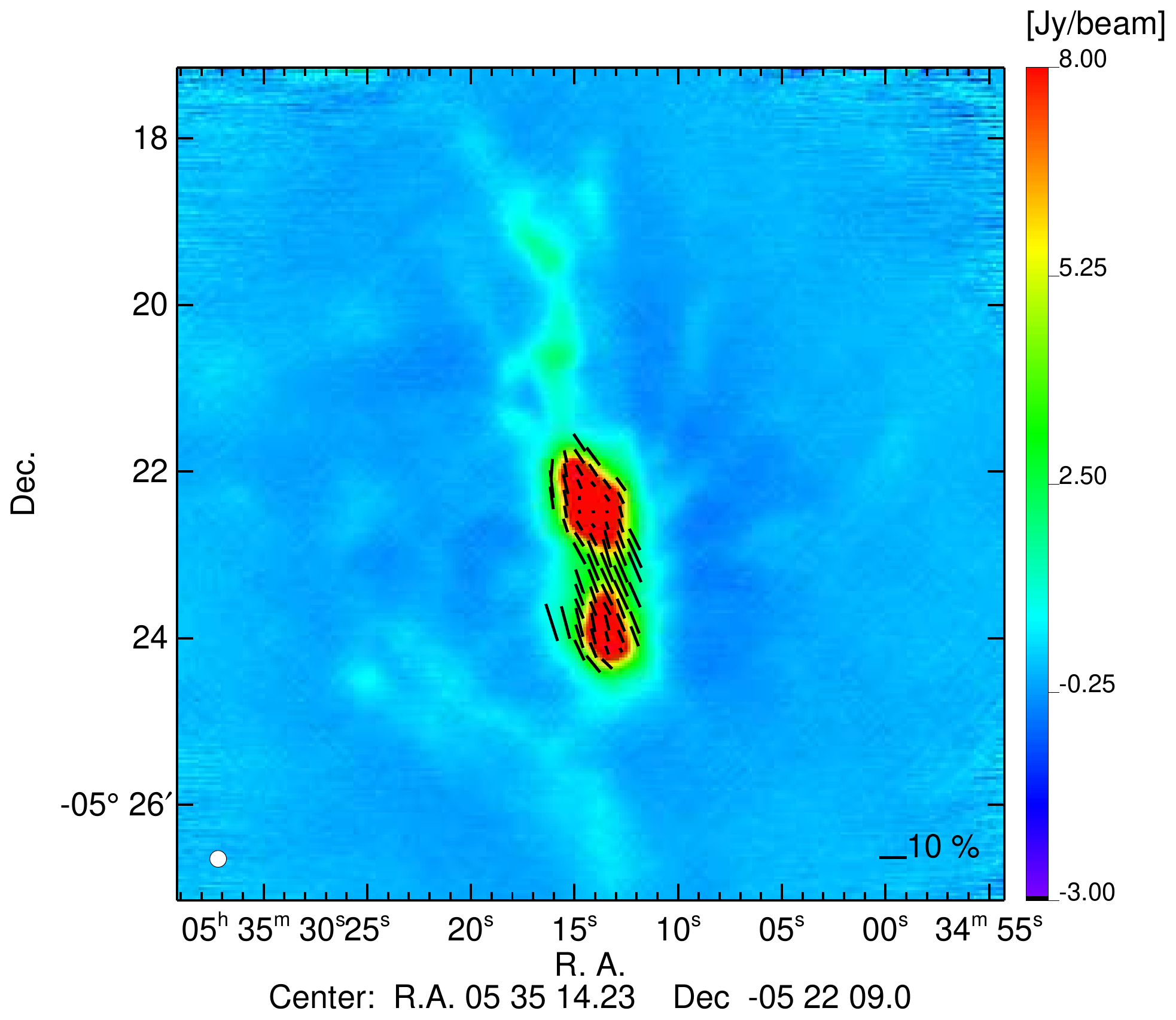}
    \includegraphics[scale=0.25]{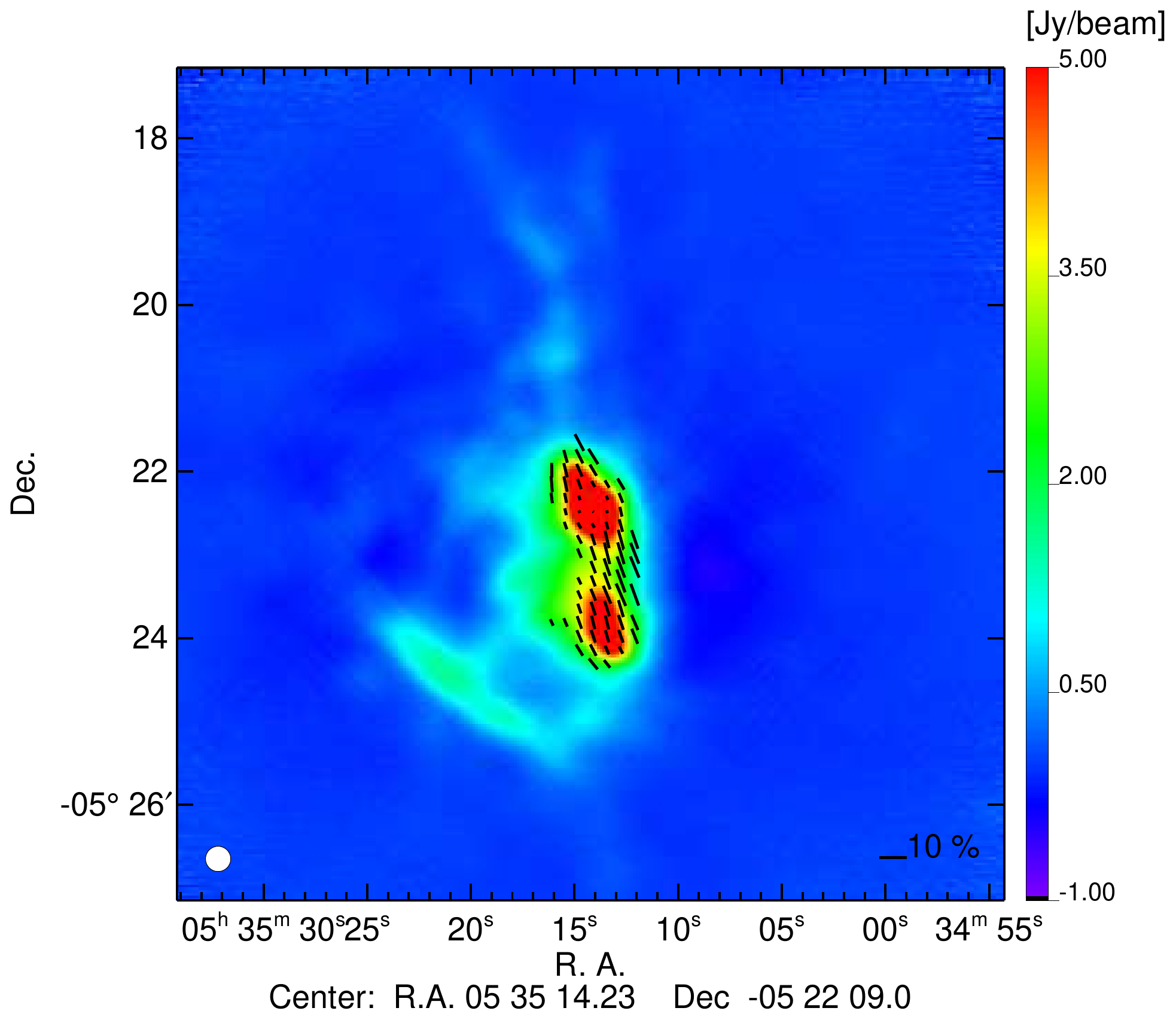}
    \label{fig1}       % Give a unique label
    \caption{Stokes I maps of Orion OMC-1 observed with the NIKA camera at 260 GHz (left) and 150 GHz (right). The polarization vectors are overplotted where the SNR for the polarization intensity $>$ 2.}
  \end{center}
\end{figure}

\begin{figure}[ht]
  \begin{center}
    %\vspace*{5cm}
  \includegraphics[scale=0.23]{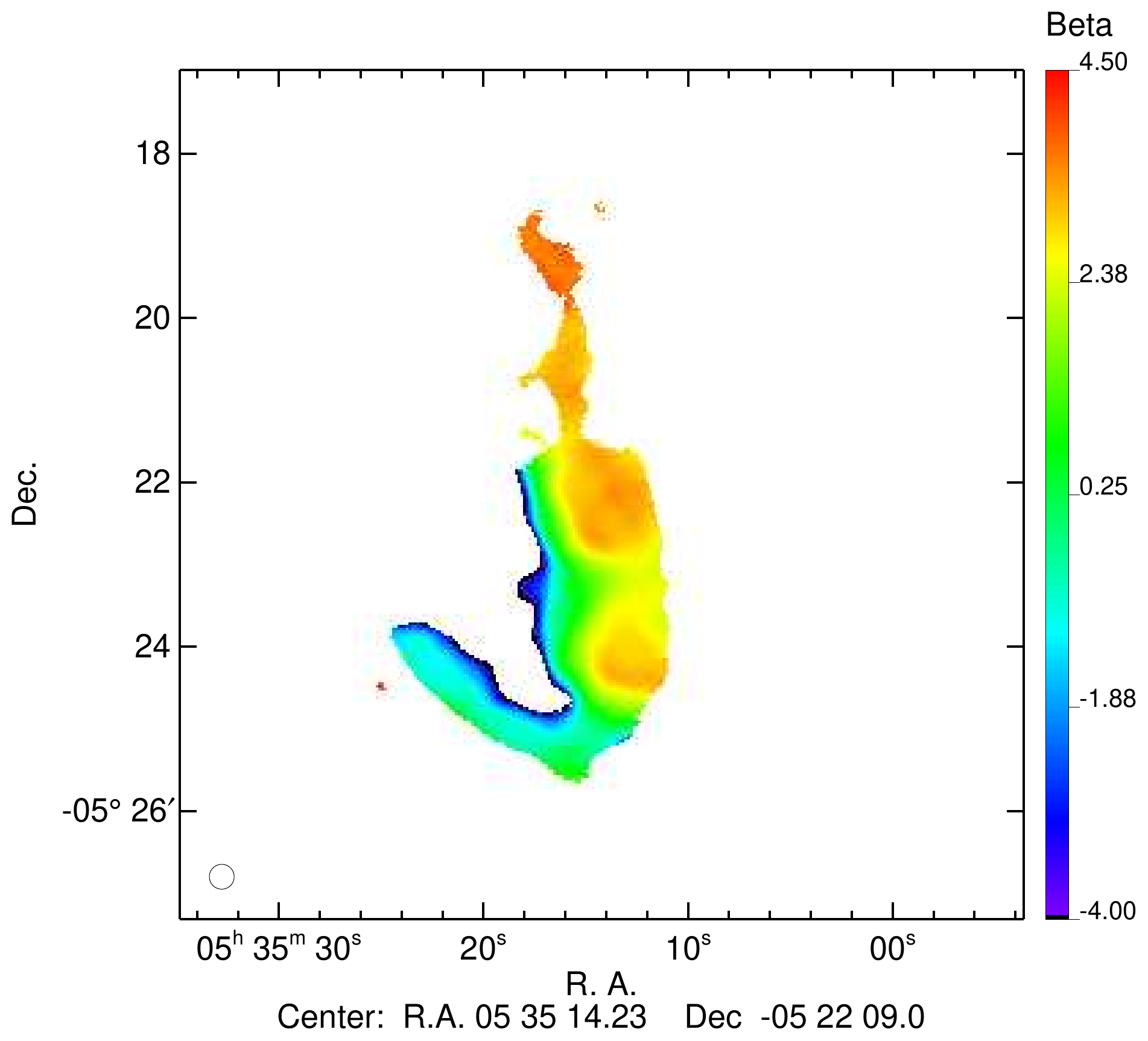}
  \includegraphics[scale=0.23]{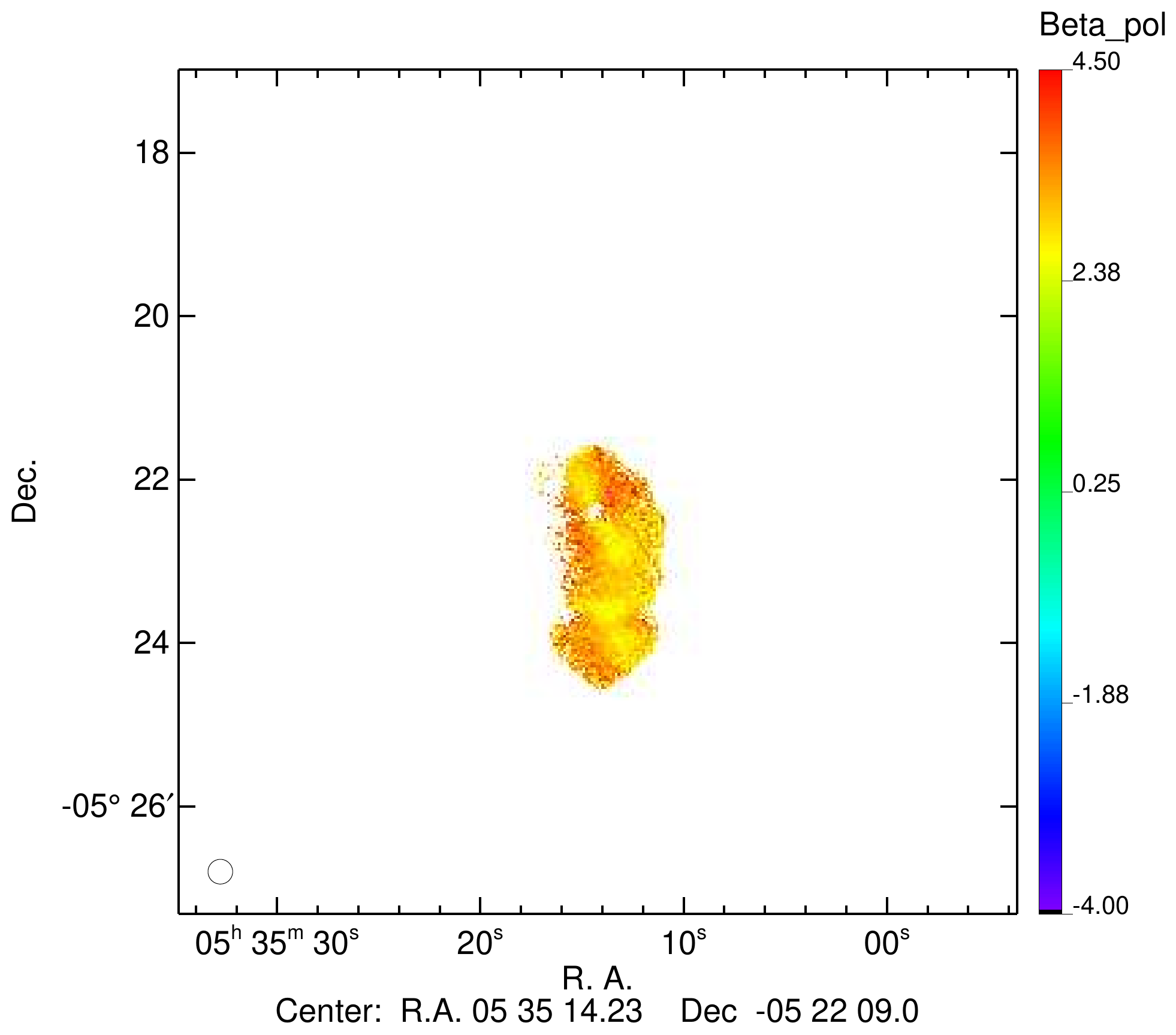}
  \includegraphics[scale=0.23]{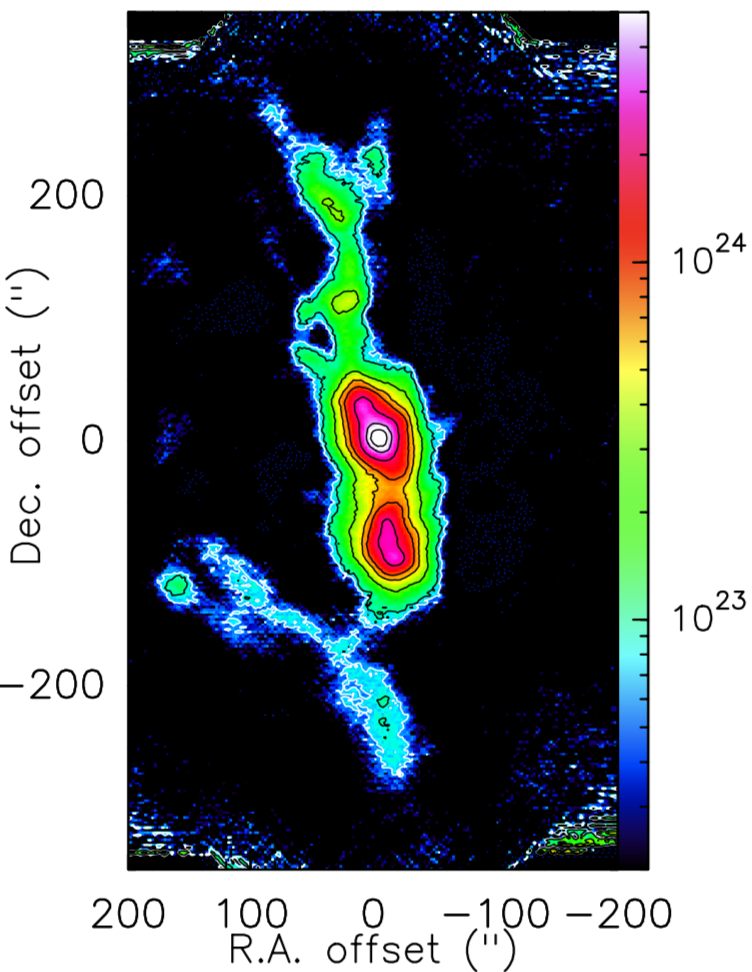}
\label{fig2}       % Give a unique label
\caption{Orion OMC-1 spectral index distribution in total intensity (left) and polarization (middle). Column density map NH$_2$ (right panel) obtained from the intensity map Stokes I at 1.15 mm.}
\end{center}
\end{figure}

\section{Crab nebula polarization observation at 150 GHz}
The Crab nebula is a supernova remnant emitting a highly polarized signal \cite{hester}. The synchrotron emission is observed in the radio frequency domain and is powered by the pulsar located at equatorial coordinates (J2000) $R.A. = 5^h34^m31.9383014s$ and $Dec. = 22^{\circ}0^{\prime}52.17577^{\prime\prime\}}$ \cite{lobanov} through its jet.
The Crab nebula is
the most intense polarized astrophysical object in the microwave
sky at angular scales of a few arcminutes and for this reason
it is chosen not only for high resolution cameras calibration, but also
for cosmic microwave background (CMB)
polarization experiments, which have beamwidths comparable
to the extension of the source of about 5$^\prime$. Upcoming CMB experiments
aiming at measuring the primordial B-modes require an accurate
determination of the foreground emissions to the CMB
signal and a high control of systematic effects.
In this section we summarize an extensive study provided by \cite{ritacco2018}, which presents the first high angular resolution polarization observations of the Crab nebula at 150 GHz performed with the NIKA camera and includes all the polarization observations available at millimeter wavelenghts to estimate the polarization spectral energy distribution (SED).

\begin{figure}[ht]
  \begin{center}
 %   \vspace*{5cm}
    \includegraphics[scale=0.25]{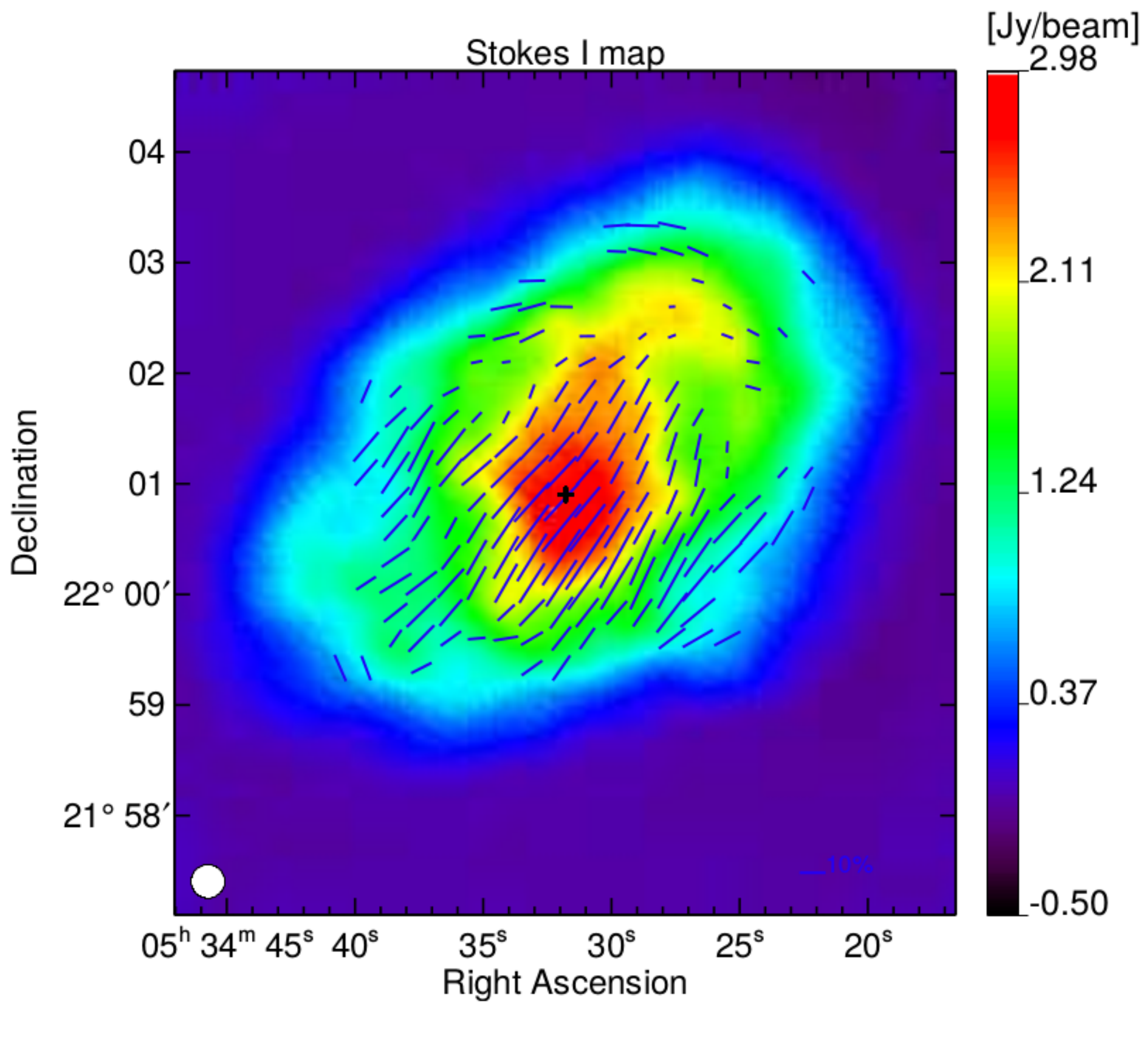}
    \includegraphics[scale=0.35]{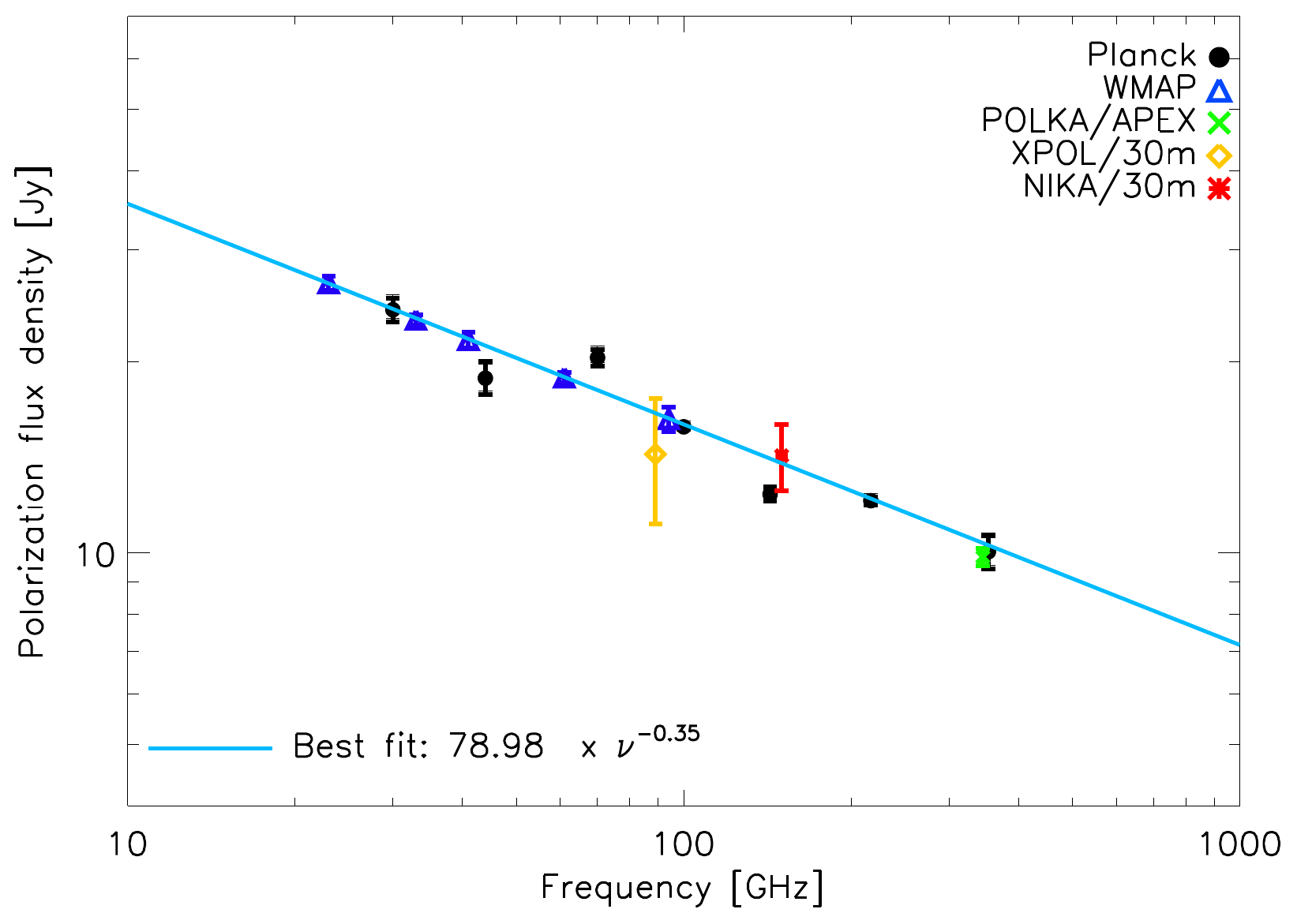}
\label{fig3}       % Give a unique label
\caption{{\it Left}: Total intensity map at 150 GHz with polarization vectors overplotted where the SNR for the polarization intensity $>$ 3. The black cross indicates the position of the pulsar. {\it Right}: Crab nebula polarization flux SED \cite{ritacco2018} and references therein.}
\end{center} 
\end{figure}

Figure~3 shows on left panel the Stokes I map of the Crab nebula as obtained with the NIKA camera at 150 GHz with polarization vectors overplotted. In Figure~3 (right frame) it is shown the spectral energy distribution in polarization obtained using all the available observations in polarization from CMB satellites and high resolution experiments. Fitting a single power law as discussed in \cite{ritacco2018} we estimated the spectral index $\beta_{pol}$ = -0.347$\pm$0.026. This result is consistent with the total power spectral index and confirms that the synchrotron emission is the fundamental mechanism that drives the polarization emission of the Crab nebula. \cite{ritacco2018} found also that the polarization angle of the Crab nebula is consistent with being constant with frequency, from 20 to 353 GHz, at arcmin scales with a value of $-87.7^{\circ}\pm0.3$ in Galactic coordinates. This result led another study centered on the analysis of the impact of such an uncertainty on the estimation of $r$ parameter (tensor-to-scalar ratio) that is directly related to the energy scale of the inflation \cite{aumont2019}. 

\section{Conclusions}
In addition to several scientific results obtained in total intensity the NIKA camera represented a test bench for polarization observations with NIKA2. During few observational campaigns the polarization observations obtained on several compact, extended and diffuse sources demonstrated the ability of such a polarization system in precisely reconstructing the polarization of the sky. In this short paper we have summarized the main results obtained on two known and very interesting astrophysical targets, Orion Molecular Cloud OMC-1 and the Crab nebula. The dual band images obtained with the NIKA camera of OMC-1 have been used to estimate the spatial distribution of the spectral index in total intensity and polarization. We observe a drastic change that is unexpected from standard dust emission models in regions where the 2 mm emission is stronger than the 1 mm one. This deserves a deeper multi-wavelenghts study that is beyond of the scope of this paper.

The 150 GHz observations of the Crab nebula shown here and published in \cite{ritacco2018} together with other experiments observations show that a single population of relativistic electrons is responsible of the synchrotron radiation observed in both total power and polarization.

These results have shown for the first time polarization observations obtained with a continuously rotating HWP coupled to a KIDs based camera. This opened the way to the incoming polarization observations of the NIKA2 camera and future experiments development that aim at using the same detection strategy for sensitive polarization observations.
%For two-column wide figures use syntax of figure~\ref{fig-2}
%\begin{figure*}
%\centering
%\vspace*{5cm}       % Give the correct figure height in cm
%\label{fig-2}       % Give a unique label
%\end{figure*}

%For tables use syntax in table~\ref{tab-1}.
%\begin{table}
%\centering
%\label{tab-1}       % Give a unique label
% For LaTeX tables you can use
%\begin{tabular}{lll}
%\hline
%first & second & third  \\\hline
%number & number & number \\\hline
%\end{tabular}
% Or use
%\vspace*{5cm}  % with the correct table height
%\end{table}

%\newpage %%to test - remove when writing your article
\section*{Acknowledgements}
We would like to thank the IRAM staff for their support during the campaigns. The NIKA dilution cryostat has been designed and built at the Institut N\'eel. In particular, we acknowledge the crucial contribution of the Cryogenics Group, and in particular Gregory Garde, Henri Rodenas, Jean Paul Leggeri, Philippe Camus. This work has been partially funded by the Foundation Nanoscience Grenoble and the LabEx FOCUS ANR-11-LABX-0013. This work is supported by the French National Research Agency under the contracts "MKIDS", "NIKA" and ANR-15-CE31-0017 and in the framework of the "Investissements d’avenir” program (ANR-15-IDEX-02). This work has benefited from the support of the European Research Council Advanced Grant ORISTARS under the European Union's Seventh Framework Programme (Grant Agreement no. 291294). F.R. acknowledges financial supports provided by NASA through SAO Award Number SV2-82023 issued by the Chandra X-Ray Observatory Center, which is operated by the Smithsonian Astrophysical Observatory for and on behalf of NASA under contract NAS8-03060.

%
% BibTeX or Biber users please use (the style is already called in the class, ensure that the "woc.bst" style is in your local directory)
% \bibliography{name or your bibliography database}

\end{document}